\documentclass[a4paper,11pt]{article}
\usepackage{pos}
\usepackage{physics}
\usepackage{braket}
\usepackage{booktabs}
\usepackage{arydshln} 
\usepackage{hhline}

\title{Inclusive and exclusive semileptonic decays of heavy mesons on the lattice}
\author*[a]{Zhi Hu}
\author[b]{Alessandro Barone}
\author[c]{Ahmed Elgaziari}
\author[d,e]{Shoji Hashimoto}
\author[c,f,g]{Andreas J\"uttner}
\author[d,e]{Takashi Kaneko}
\author[a]{Ryan Kellermann}

\affiliation[a]{High Energy Accelerator Research Organization (KEK), Ibaraki 305-0801, Japan}

\affiliation[b]{PRISMA+ Cluster of Excellence \& Institut f\"ur Kernphysik, Johannes-Gutenberg-Universit\"at Mainz, D-55099 Mainz, Germany}

\affiliation[c]{School of Physics and Astronomy, University of Southampton, Southampton SO17 1BJ, UK}

\affiliation[d]{School of High Energy Accelerator Science, SOKENDAI (The Graduate University for Advanced Studies), Ibaraki 305-0801, Japan}

\affiliation[e]{High Energy Accelerator Research Organization (KEK), Ibaraki 305-0801, Japan}

\affiliation[f]{STAG Research Center, University of Southampton, Southampton SO17 1BJ, UK}

\affiliation[g]{CERN, Theoretical Physics Department, Geneva, Switzerland}

\emailAdd{huzhi@post.kek.jp}
\emailAdd{abarone@uni-mainz.de}
\emailAdd{A.Elgaziari@soton.ac.uk}
\emailAdd{shoji.hashimoto@kek.jp}
\emailAdd{andreas.juttner@cern.ch}
\emailAdd{takashi.kaneko@kek.jp}
\emailAdd{kelry@post.kek.jp}

\abstract{We report the recent progress from our group in extracting observables of both inclusive and exclusive semileptonic heavy-meson decays directly from lattice QCD four-point correlators. On the inclusive side, we illustrate how to estimate the systematic uncertainties from omitted higher-order terms and non-zero smearing of the kernel approximation, building on two important features of the Chebyshev expansion. On the exclusive side, we perform BCL parameterizations of the pseudoscalar to pseudoscalar form factors and compare the fitted coefficients with those from earlier results by HPQCD. We also perform a HQET-based parameterization of the P-wave form factors to shed new light on the 1/2-vs-3/2 puzzle. This work constitutes a step toward a unified lattice treatment of inclusive and exclusive semileptonic decays, relevant for the $\Abs{V_{cb}}$ puzzle. In this study, we use lattice ensembles from the RBC/UKQCD collaboration for numerical investigations. Future developments from our group will focus on the control of other systematic effects for inclusive decays and investigations of other techniques with reduced statistical errors to extract exclusive contributions from lattice four-point correlators.}

\FullConference{The 17th International Conference on Heavy Quarks and Leptons (HQL 2025)\\
15-19 September, 2025\\
Peking University, Beijing, China\\}

\newcommand{\ie}{\textit{i.e.}} 
 
\newcommand{\fig}[1]{Fig.~#1}

\newcommand{\eq}[1]{Eq.~(#1)}

\newcommand{\paper}[1]{Ref.~#1}
\newcommand{\papers}[1]{Refs.~#1}

\newcommand{\mybf}[1]{\boldsymbol{#1}}

\newcommand{\rmd}{\mathrm{d}}
\newcommand{\MBs}{M_{B_s}}

\newcommand{\Abs}[1]{\left|#1\right|}
\newcommand{\src}{\mathrm{src}}
\newcommand{\snk}{\mathrm{snk}}

\begin{document}
\maketitle

\section{Introduction}

A precise determination of the CKM matrix element $\Abs{V_{cb}}$ requires continuous interplay between theory and experiment. The golden channel to measure $\Abs{V_{cb}}$ is the semileptonic decay of
$
    B_{(s)}\rightarrow X_c(X_{cs}) l \nu_l 
$, for which experimental data for both inclusive (decays into all possible $X_c(X_{cs})$) and exclusive (decays into one specific $X_c(X_{cs})$) modes exist~\cite{Belle-II:2023okj,Belle-II:2022evt}. Traditionally, inclusive and exclusive semileptonic decays are treated within different theoretical frameworks: inclusive observables are computed using the operator product expansion (OPE)~\cite{Bordone:2021oof,Bernlochner:2022ucr}, while exclusive modes rely mainly on form factors extracted from lattice three-point correlators~\cite{FermilabLattice:2021cdg,Harrison:2023dzh,Aoki:2023qpa}. Despite substantial improvements on the theoretical calculations of both modes~\cite{Gambino:2016jkc,Fael:2018vsp,Bourrely:2008za,Bigi:2017jbd}, the long-standing tension between inclusive and exclusive determinations of $\Abs{V_{cb}}$ persists~\cite{FlavourLatticeAveragingGroupFLAG:2024oxs,HeavyFlavorAveragingGroupHFLAV:2024ctg}.

A possible source of this discrepancy is the presence of unknown systematic effects introduced by the use of different theoretical formalisms for inclusive and exclusive modes. This motivates developing new theoretical approaches that could treat both modes within a unified framework, enabling a more direct investigation of the $\Abs{V_{cb}}$ puzzle. 

%In this proceeding, we report recent methodological developments from our group in extracting observables of both inclusive and exclusive semileptonic decays directly from lattice four-point correlators~\cite{Hashimoto:2017wqo}. \marginpar{\rk{It might be helpful for the reader to ''define'' four-point functions. In the sense of that they are the tools that allow us to treat inclusive decays. You do that in the following paragraph, so you might want to just remove this last sentence completely.}}

On the one hand, the calculations of inclusive observables have long been a challenging task for lattice quantum chromodynamics (QCD). It can be easily noticed that hadronic tensors $W_{\mu\nu}$, which are the non-perturbative parts of most inclusive observables, are also the spectral representations of the four-point correlators $C_{J_\mu J_\nu}$, which are calculable on the discrete Euclidean lattice, \ie{}
\begin{gather}
    C_{J_\mu J_\nu}(\mybf{q} , t) \equiv \int \rmd^3 \mybf{x}\frac{ e^{i\mybf{q}\cdot\mybf{x}}}{2M_H} \Braket{H| J_\mu^\dagger(\mybf{x},0) e^{-\hat{H}t} J_\nu(0) |H} = \int_{0}^{\infty} \rmd E_X e^{-tE_X} W_{\mu\nu}(\mybf{q} , E_X) \, .\label{C_munu_def} 
\end{gather}
Here, $q=(q^0,\mybf{q})$ is the momentum of the lepton pair, $E_X$ is the energy of the final-state hadronic system, and the inserted current, $J_\mu$ and $J_\nu$, can be either vector or axial. We work in the center-of-mass frame of the initial meson $H$ with $p_H = (M_H,\mybf{0})$. Nevertheless, the inverse Laplace transform, when performed numerically, is highly unstable, making the extraction of hadronic tensors from lattice four-point correlators an infamous inverse problem~\cite{Regularization}. To circumvent this obstacle, recent methodological developments have demonstrated that the final-state energy integration allows inclusive observables to be reconstructed directly from four-point correlators~\cite{Hashimoto:2017wqo,Barone:2023tbl,Kellermann:2025pzt,Hansen:2019idp,DeSantis:2025yfm,DeSantis:2025qbb}. However, the control of systematic effects in inclusive calculations remains to be carefully developed and discussed. In this proceeding, we mainly report on our developments towards better controling the systematic effects resulting from kernel approximations.

On the other hand, since four-point correlators do not require explicit interpolating operators for final states and naturally encode contributions from all decay channels, they offer improved access to the decays into excited final states and may also shed new light on the so-called $1/2$-vs-$3/2$ puzzle~\cite{Bigi:2007qp}. Although existing studies remain at the proof-of-concept level~\cite{Hu:2025hpn,Bailas:2019diq}, systematic isolation of exclusive contributions appears achievable through multi-exponential fitting. In this report, we also present form factors extracted from lattice four-point correlators in certain interesting channels and discuss their phenomenological implications.

\section{Lattice setup}

% Based on our analysis in \paper{\cite{Kellermann:2025pzt}}, we use the $D_{s}\rightarrow X_{s\bar{s}}$ semileptonic decay corresponding to $V_{cs}$ to demonstrate our method to control the systematic effects of inclusive decays. The same formalism can be straightforwardly generalized to investigating $B_{(s)}\rightarrow X_c(X_{cs})$ decays and $\Abs{V_{cb}}$. For this, we employ a gauge ensemble generated by the JLQCD Collaboration~\cite{Colquhoun:2022atw,Kellermann:2025pzt}, with M\"obius domain-wall fermion (DWF) action for both sea (2+1 flavours) and valence (light, strange and charm) quarks. Simulations are carried out on a $48^3\times 96$ lattice with lattice spacing $a^{-1}\approx 3.610(9)\,\mathrm{GeV}$ and light-quark masses corresponding to $M_{\pi}\approx 300\,\mathrm{MeV}$.

For numerical illustrations, we focus on $B_{s}\rightarrow X_{cs}$ decays. We use two $ 24^3\times 64 $ lattices generated by the RBC/UKQCD Collaboration~\cite{Flynn:2023nhi,Barone:2023tbl}, with $a\approx 0.11\,\mathrm{fm}$. Light (2+1 flavours), $c$ and $b$ quarks are simulated using the domain-wall fermion (DWF), M\"obius DWF, and relativistic-heavy-quark action, respectively. The choices of light quark masses in both ensembles correspond to $M_{\pi}\approx 330\,\mathrm{MeV}$. The four-point correlators are simulated on the lattice with $B_s$-meson interpolators at source and sink, and two $V-A$ currents inserted in between
\begin{gather}
    C_{J_\mu J_\nu}(\mybf{q},t=t_2 - t_1) \propto \int \rmd^3 \mybf{x} \; e^{i\mybf{q}\cdot\mybf{x}} \Braket{0| \phi_{B_s}({t_{\snk}}) J^{\dagger}_\mu(\mybf{x},t_2) J_\nu(\mybf{0},t_1) \phi_{B_s}^{\dagger}({t_{\src}}) |0} \, . \label{eq:def_four_pt}
\end{gather}
We fix $ t_\snk - t_\src = 20 $, $ t_2 - t_\src = 14 $, with $ t_\src \leq t_1\leq t_2 $. Here and in what follows we set the lattice spacing $a=1$ to simplify the formulas.

\section{Inclusive decays}

Observables of inclusive semileptonic decays can be schematically expressed as
\begin{gather}
    \bar{X}(\mybf{q}^2) \equiv \frac{\rmd \Gamma^{\mathrm{inc}}}{\rmd \mybf{q}^2} = \left|V_{fg}\right|^2 \int_{E_X^{\mathrm{min}}}^{E_X^{\mathrm{max}}} \rmd E_X W_{\mu\nu} (\mybf{q}^2, E_X) k^{\mu\nu}(\mybf{q}^2, E_X) \, ,
\end{gather}
where $k^{\mu\nu}$ contains some known kinematical factors, and $V_{fg}$ is the relevant CKM matrix element. To exploit the spectral representation \eq{\ref{C_munu_def}}, we have to extend the integral interval to $[0,\infty)$. Since the spectral representation only starts at $E_X^{\mathrm{min}}$, one can choose the lower limit arbitrarily as long as $E_X^0<E_X^{\mathrm{min}}$. However, it is necessary to introduce a step function $\theta(E_X^{\mathrm{max}} - E_X)$ to cut off contributions above the kinematical region of semileptonic decays. To regulate the singular behaviour of the step function, during the numerical implementation, we further replace it with a sigmoid function with the smearing parameter $\sigma$. Finally, we have
$
    \bar{X}_\sigma (\mybf{q}^2) = \int_{E_X^{0}}^{\infty} \rmd E_X W_{\mu\nu} (\mybf{q}^2, E_X) K_\sigma^{\mu\nu}(\mybf{q}^2, E_X) \, ,
$
with $K_\sigma^{\mu\nu}\equiv k^{\mu\nu} \theta_\sigma(E_X^{\mathrm{max}} - E_X)$. If the kernel $K_\sigma^{\mu\nu}$ can be approximated as $K_\sigma^{\mu\nu}\approx \sum_{k}^{N} a_{\sigma,k}^{\mu\nu} e^{-E_X k}$, then inclusive observables can be obtained directly from lattice four-point correlators as
\begin{gather}
    \bar{X}_\sigma (\mybf{q}^2) \approx \sum_{k}^{N} a_{\sigma,k}^{\mu\nu} C_{J_\mu J_\nu}(\mybf{q}^2,t=k) \, . \label{eq:aj}
\end{gather}

The calculations of $\bar{X}$ are thus split into two parts: estimations of the approximation coefficients $a_{\sigma,k}^{\mu\nu}$, and simulations of $C_{J_\mu J_\nu}$ on a lattice, which are subject to different sources of systematic uncertainties. In this proceeding, we only focus on the effects of $\sigma\rightarrow 0$ and $N\rightarrow \infty$ limits for $a_{\sigma,k}^{\mu\nu}$. Interested readers may find further discussions in \paper{\cite{Kellermann:2025pzt}} and our upcoming papers. Noticing that both $\sigma\rightarrow 0$ and $N\rightarrow \infty$ limits correspond to including more and more high-frequency components in the kernel approximations, it is only natural to perform these two limits simultaneously. Here, we set $\sigma = 1/N$. However, since the highest accessible order in the exponential approximation is constrained by $t_2-t_{\mathrm{src}}$ in the lattice simulations, these two limits cannot be taken exactly and must be estimated.

\begin{figure}
    \centering
    \includegraphics[width=0.9\linewidth]{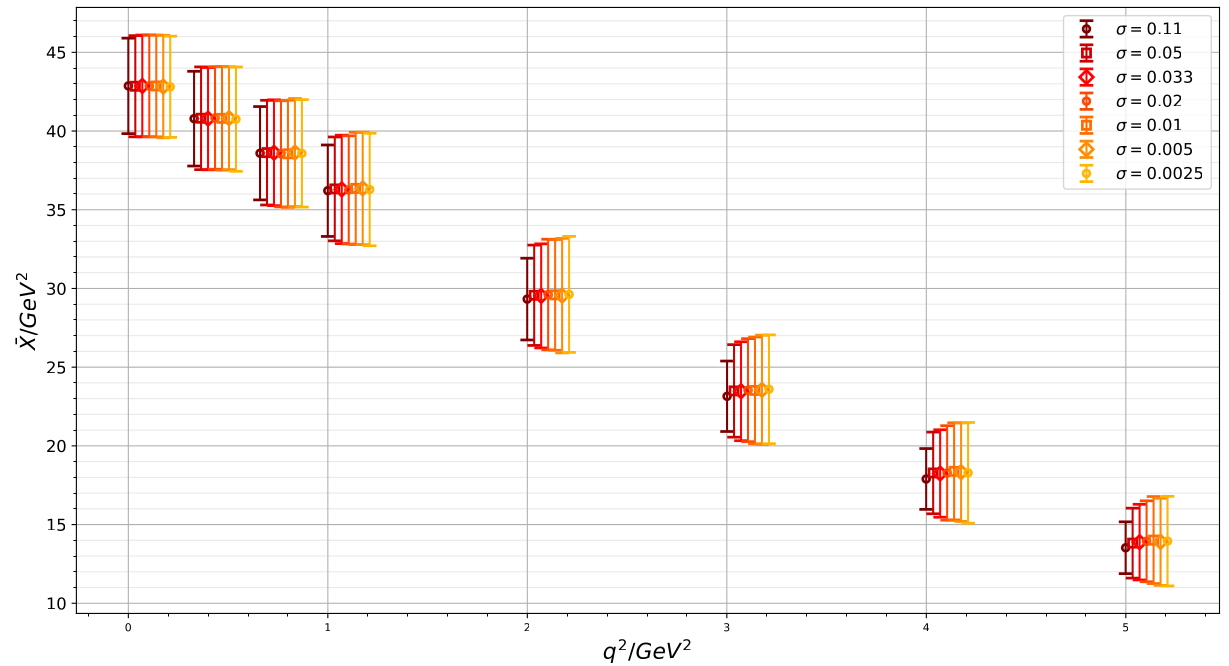}
    \caption{We present the effects of the $N\rightarrow\infty$ and $\sigma\rightarrow 0$ limits, taken simultaneously with $\sigma = 1/N$, on the total $\bar{X}$ of $B_s\rightarrow X_{cs}$ at all simulated momenta.}
    \label{fig:inclusive}
\end{figure}

We employ the Chebyshev method~\cite{Hashimoto:2017wqo,Barone:2023tbl,Kellermann:2025pzt} to estimate $a_{\sigma,k}^{\mu\nu}$. Readers who are interested in another systematic formalism for kernel approximation, the HLT method, may find detailed discussions in \papers{\cite{Hansen:2019idp,DeSantis:2025yfm,DeSantis:2025qbb}}. $K_\sigma^{\mu\nu}$ is expanded in (shifted) Chebyshev polynomials of the energy exponential $\tilde{T}_j(E_X)$. Chebyshev polynomials are bounded between $[-1,1]$ and the corresponding expansion coefficients $ c_j $ (which are linear combinations of coefficients $a_j$ in \eq{\ref{eq:aj}}, see more details in the Appendix of \paper{\cite{Barone:2023tbl}}) for a well-behaved function decay exponentially with respect to the order $j$~\cite{ChebyshevBook}. Building on this, we could write
\begin{gather}
    \bar{X}_\sigma = \sum_{j=0}^N c_{\sigma,j}^{\mu\nu} \Braket{\tilde{T}_j}_{\mu\nu} = \sum_{j=0}^{N_{\mathrm{cut}}} c_{\sigma,j}^{\mu\nu} \Braket{\tilde{T}_j}_{\mu\nu} + \sum_{j=N_{\mathrm{cut}}+1}^{N} c_{\sigma,j}^{\mu\nu} \Braket{\tilde{T}_j}_{\mu\nu} \, ,
\end{gather}
where $N_{\mathrm{cut}}$ is the largest order allowed by $t_2-t_{\mathrm{src}}$ and $\Braket{\tilde{T}_j}_{\mu\nu}\equiv\int_{E_X^{0}}^{\infty} \rmd E_X W_{\mu\nu} (\mybf{q}^2, E_X) \tilde{T}_j$. While we do not have direct access to the second term from lattice simulations, we could estimate its largest possible contribution as systematic uncertainty 
\begin{gather}
    \delta^2_{\mathrm{max}} \sim \sum_{j=N_{\mathrm{cut}}+1}^{N} \left|c_{\sigma,j}^{\mu\nu}\right|^2 \, . 
\end{gather}
To better control systematic effects, we also extract ground-state contributions to $C_{J_\mu J_\nu}$ by multi-exponential fitting and calculate the corresponding parts of $\bar{X}$ separately. The Chebyshev approximation and estimation of the systematic effects are thus only applied to the remaining $C_{J_\mu J_\nu}$ after subtracting the ground-state contributions. In \fig{\ref{fig:inclusive}}, we demonstrate the effects of pushing $\sigma=1/N=1/9$ to $\sigma=1/N=1/400$ for the total (resumming ground-state and excited-state contributions and summing over all contributing currents) $\bar{X}$ of $B_s\rightarrow X_{cs}$ at all simulated momenta. We notice that larger momentum leads to larger dependence on $\sigma=1/N$, which is due to the shrinking of the phase space. Nevertheless, both the central values and the uncertainties in \fig{\ref{fig:inclusive}} converge very fast with respect to $\sigma=1/N$ at all momenta, which results from the separation of the ground-state and excited-state contributions, and the exponential decay of the expansion coefficients. Whether these treatments yield similar stability for other kernels, such as those for moments of inclusive decays, requires further numerical verification.

% we demonstrate the effects of pushing $\sigma=1/N=1/9$ to $\sigma=1/N=1/400$ \hz{for the total (summing ground-state and excited-state contributions and summing over all contributing currents) $\bar{X}$ of $B_s\rightarrow X_{cs}$ at the smallest simulated momentum $\mybf{q}^2 = 0.33\,\mathrm{GeV^2}$. The quick convergences of both the central values and the uncertainties result from the separation of the ground-state and excited-state contributions and the exponential decay of the expansion coefficients. Whether these treatments lead to similar statistical stability for other kernels, like those for moments of inclusive decays, needs further numerical checks.}

\section{Exclusive decays}

By inserting a complete basis of the final states between the two currents, four-point correlators admit a multi-exponential representation,
\begin{gather}
    C_{J_\mu J_\nu}(\mybf{q}^2,t) = \sum_{X_{cs}} \frac{1}{2E_{X_{cs}} 2 M_{B_s}} \Braket{B_s| J_\mu^\dagger(0)|X_{cs}} \Braket{ X_{cs} | J_\nu(0) |B_s} e^{-E_{X_{cs}} t} \, .
\end{gather}
Multi-exponential fits to the lattice data, therefore, allow the extraction of transition form factors of different decay channels, which are Lorentz-invariant parameterizations of $\Braket{ X_{cs} | J_\nu(0) |B_s}$. The spectrum of $X_{cs}$ can be organized as doublets with different radial and orbital angular momentum quantum numbers, $(n)$ and $L$. Here we focus on the radial ground state, $(n)=(0)$. With $L=0$, we have the $S$-wave doublet $\left(D_s, D_s^*\right)$. With $L=1$, we have one $P_{1/2}$ doublet $\left(D_{s0}^* , D_{s1}^\prime\right)$ and one $P_{3/2}$ doublet $\left(D_{s1} , D_{s2}\right)$. Explicit formulas to connect their form factors and $ C_{J_\mu J_\nu} $ can be found in \papers{\cite{Hu:2025hpn,Bailas:2019diq}}.

\begin{figure}
    \centering
    \includegraphics[width=0.8\linewidth]{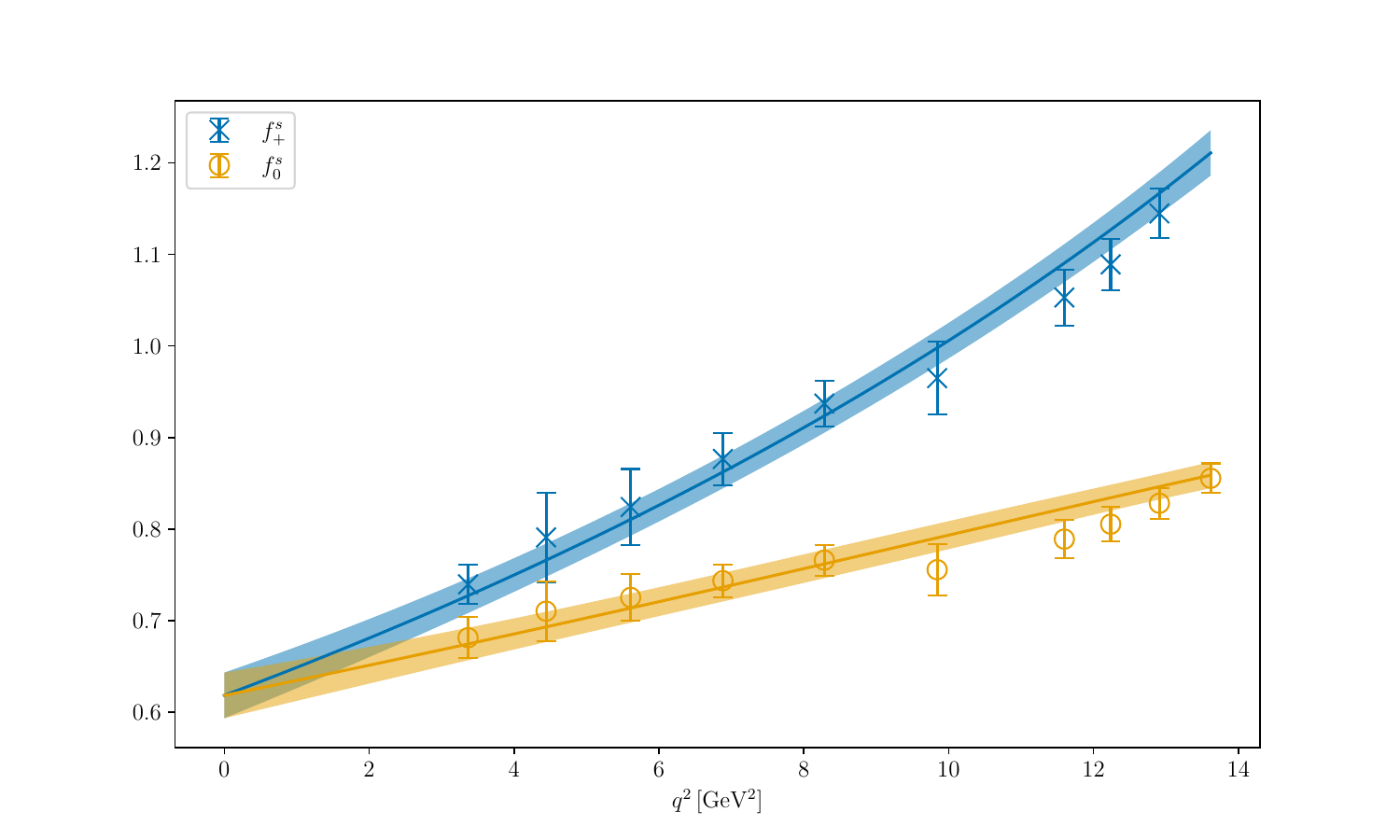}
    \caption{The BCL parameterisations (lines) are compared with the lattice data (data points) for pseudoscalar to pseudoscalar form factors for $B_s$ semileptonic decays, \ie{}, $f_+^s$ and $f_0^s$. Noticing that we have a somewhat different $q^2$ range compared with that in \paper{\cite{McLean:2019qcx}} due to unphysical choices of heavy quark masses in our simulations.}
    \label{fig:exclusive}
\end{figure}

We first focus on the extracted form factors of $B_s \rightarrow D_s$, $f_+^s$ and $f_0^s$. Our simulations are performed with ten different $\mybf{q}^2$ values spanning the full kinematical range allowed by this semileptonic decay. Their $\mybf{q}^2$-dependence is described by a modified Bourrely-Caprini-Lellouch (BCL) parameterization from HPQCD 19~\cite{McLean:2019qcx}:
% \begin{align}
%     f_+^s(q^2) &= \frac{1}{P_+} \sum_{n=0}^2 a_n^+\left[ z^n(q^2) - \frac{n}{3} (-1)^{n-3}z^3(q^2) \right] \, , \label{eq:BCL_fPlus}\\
%     f_0^s(q^2) &= \frac{1}{P_0} \sum_{n=0}^2 a_n^0 z^n(q^2) \, . \label{eq:BCL_fZero}
% \end{align}
$
    f_+^s(q^2) = \frac{1}{P_+} \sum_{n=0}^2 a_n^+\left[ z^n(q^2) - \frac{n}{3} (-1)^{n-3}z^3(q^2) \right] \, ,
    f_0^s(q^2) = \frac{1}{P_0} \sum_{n=0}^2 a_n^0 z^n(q^2) \, .
$
Blaschke factors $P_{0,+}$ are included to account for poles above the semileptonic region. The pole masses differ from those used in HPQCD 19~\cite{McLean:2019qcx} in order to accommodate the unphysical light-quark masses in our simulations. The variable $z$ is defined as $z(q^2) \equiv \frac{\sqrt{t_+ - q^2} - \sqrt{t_+}}{\sqrt{t_+ - q^2} + \sqrt{t_+}}$, where $t_+ \equiv (\MBs + M_{D_s})^2$. With these choices, the kinematical constraint $f_+^s(0) = f_0^s(0)$ reduces to a simple relation $a_0^+ = a_0^0$. In total, five BCL coefficients are fitted to describe the full $q^2$-dependence of both $f_+^s$ and $f_0^s$. The resulting fits are shown in \fig{\ref{fig:exclusive}}. We have two observations compared with results from \paper{\cite{McLean:2019qcx}}. First, the values of the zeroth- and first-order coefficients are consistent between the two studies, despite the use of different lattice actions. Our results have larger uncertainties, which is expected, as this proof-of-concept study relies on a single ensemble and does not include any physical extrapolations. Second, the third-order coefficients are dominated by statistical uncertainties and thus can not be effectively determined in both calculations. This is not surprising given that $\abs{z}<0.06$ for this decay mode.

We now turn to the $P$-wave form factors. In our analysis, only the contributions from $B_s\rightarrow D_{s0}^*$ for the $P_{1/2}$ wave and $B_s\rightarrow D_{s1}$ for the $P_{3/2}$ wave are kept (see \papers{\cite{Hu:2025hpn,Bailas:2019diq}} for detailed arguments). These decays are described by six form factors: $g_+^s$, $g_-^s$, $f_{V1}^s$, $f_{V2}^s$, $f_{V3}^s$ and $f_{A}^s$. We then follow the approximation $C$ from \papers{\cite{Leibovich:1997em,Bernlochner:2016bci}} to parameterise their dependence on $w\equiv E_X / M_X = \sqrt{1+\mybf{q}^2/M_X^2}$, which is based on the following assumptions: 1) the heavy-quark expansions are truncated to the first order in the inverse heavy-quark masses; 2) the terms from chromomagnetic and kinematic energy operators are ignored; 3) the terms from matching HQET currents to QCD currents, $ \tau_1 , \tau_2 , \zeta_1 $, are assumed to be proportional to the leading Isgur-Wise functions $ \tau,\zeta $; 4) the leading Isgur-Wise functions are expanded linearly with respect to $w$, $\tau(w) = \sqrt{3} \tau_{3/2}^{(0)}\left[ 1 + \tau^\prime (w-1) \right]$ and $\zeta(w) = 2 \tau_{1/2}^{(0)} \left[ 1 + \zeta^\prime (w-1) \right]$, where the superscript ${(0)}$ denotes radial ground states. In this way, the fit involves in total of seven parameters from the heavy-quark effective theory (HQET) $\{ \tau_{3/2}^{(0)},\tau^\prime,\tau_{1/2}^{(0)}, \zeta^\prime , \hat{\tau}^1, \hat{\tau}^2, \hat{\zeta}^1\}$. From the fit we obtain $\left|\tau_{1/2}^{(0)}\right|^2 - \left|\tau_{3/2}^{(0)}\right|^2 = 0.021\pm 0.076$. Including the full $w$-dependence shifts the central value compared to our earlier estimate based solely on zero-recoil form factors~\cite{Hu:2025hpn}. However, both results are consistent with zero, and one still expects non-negligible contributions from radial excited states to the Uraltsev sum rule~\cite{Uraltsev:2000ce}. The extracted slopes, $\tau^\prime = -1.94\pm 0.85$ and $\zeta^\prime = -0.4\pm 2.3$, indicate that the $w$-dependence is not necessarily mild. Consequently, sum rules evaluated at the kinematical endpoint may not be straightforwardly extended to the full kinematical region and further dedicate the proportion of the corresponding branching ratios. Together, these observations from our lattice study may offer new insight into the long-standing 1/2-vs-3/2 puzzle~\cite{Bigi:2007qp}.

% \begin{table}
%     \centering
%     \begin{tabular}{c c c c c}
%          \toprule
%          $a^0_1$ & $a_2^0$ & $a^+_0 = a_0^0$ & $a^+_1$ & $a^+_2$ \\\midrule
%          -0.24248 & -9.61441 & 0.61836 & -2.58817 & -10.4371 \\\midrule
%          0.69609 & 6.20441 & 0.01669 & 0.65081 & 5.66506 \\
%          6.20441 & 58.6077 & 0.12821 & 5.84714 & 54.5366 \\
%          0.01669 & 0.12821 & 0.00062 & 0.01541 & 0.11655 \\
%          0.65081 & 5.84714 & 0.01541 & 0.66499 & 6.11029 \\
%          5.66506 & 54.5366 & 0.11655 & 6.11029 & 63.8852 \\\bottomrule
%     \end{tabular}
%     \caption{Fitted values of the BCL coefficients in \eqs{\ref{eq:BCL_fPlus}, \ref{eq:BCL_fZero}}. The first row is the central values, followed by the covariance matrix in the rest of the table.}
%     \label{tab:Ds_fPlus_fZero_BCL_parameter}
% \end{table}

\section*{Acknowledgments}
The numerical calculations of the RBC/UKQCD Collaboration use the DiRAC Extreme Scaling service at the University of Edinburgh, operated by the Edinburgh Parallel Computing Centre on behalf of the STFC DiRAC HPC Facility (www.dirac.ac.uk). DiRAC is part of the National e-Infrastructure. This equipment was funded by BEIS capital funding via STFC capital grant ST/R00238X/1 and STFC DiRAC Operations grant ST/R001006/1. The works of S.H., R.K. and T.K. are supported in part by JSPS KAKENHI Grant Numbers 22H00138, 22K21347, 23K20846 and 25K01007, and by the Post-K and Fugaku supercomputer project through the Joint Institute for Computational Fundamental Science (JICFuS). T.K. is also supported by the U.S.-Japan Science and Technology Cooperation Program in High Energy Physics (Project ID: 2024-40-2). A.J. is supported by the Eric \& Wendy Schmidt Fund for Strategic Innovation (grant agreement SIF-2023-004).

\end{document}